\documentclass[9pt,twocolumn,twoside]{opticajnl}
\journal{opticajournal} 

\setboolean{shortarticle}{true}

\usepackage{lineno}

\title{Low-cost passive single-shot ultrafast imaging at 685\,Gfps}

\author[ ]{Dilem Eşlik}
\author[ ]{Bahadır Utku Kesgin}
\author[ *]{Uğur Teğin}

\affil[ ]{Department of Electrical and Electronics Engineering, Koç University, Istanbul, Türkiye}

\affil[*]{utegin@ku.edu.tr}

\begin{abstract}
Capturing ultrafast transient phenomena conventionally requires streak cameras or computational imaging based on compressed sensing, which lead to complex and costly systems. In this Letter, we demonstrate, to the best of our knowledge, the first fully passive single-shot ultrafast imaging architecture assembled entirely from off-the-shelf, low-cost components. A commercial microlens array combined with a stack of standard microscope cover glasses maps temporal information into multiple spatial channels, and a consumer-grade CMOS image sensor records all delayed replicas within a single camera exposure. The proposed system has a total hardware cost below US\$500 and captures the evolution of a picosecond laser pulse with a temporal sampling interval of 1.46~ps, an effective frame rate of 685~Gfps, and a sequence depth of ten frames. The temporal fidelity of the system is verified by recovering the expected Gaussian pulse profile, and the spatial resolution is characterized through a point-source measurement with a point spread function of 1.86 and 1.62 pixels full width at half maximum along the horizontal and vertical directions, respectively. The proposed architecture presents an alternative approach to single-shot ultrafast imaging with a simple, low-cost, computation-free, and fully passive design.
\end{abstract}

\setboolean{displaycopyright}{false} 

\begin{document}

\maketitle

Ultrafast imaging of transient optical phenomena is important for understanding a wide range of physical processes, including laser pulse propagation, nonlinear light-matter interactions, plasma formation, fluorescence decay, and the direct visualization of light propagation itself~\cite{Velten2013Femto}. Many of these processes occur on picosecond or femtosecond timescales, and require imaging systems with frame rates well beyond the readout limits of conventional electronic detectors~\cite{abramson1978light}. Over the years, various single-shot ultrafast imaging techniques have been developed to overcome these readout limitations by converting temporal information into spatial, spectral, or frequency-domain observables within the integration time of a conventional camera~\cite{li2010frequency}.

Compressed ultrafast photography (CUP) combines a streak camera with compressed sensing reconstruction and achieves real-time imaging at frame rates exceeding $10^{11}$ frames per second~\cite{Gao2014CUP,Liang2018Review}. Later, compressed ultrafast spectral-temporal photography extended this concept using spectral encoding and pulse shearing, and reached tens of trillions of frames per second~\cite{Wang2020CUSP,Lu2019CUST}. Reconstruction constraints and optimization strategies were also reported to improve image fidelity and noise robustness in compressed ultrafast photography~\cite{zhu2016space}. In parallel, time-stretch imaging based on dispersive Fourier transformation has enabled continuous ultrafast recording at megahertz acquisition rates~\cite{Goda2009STEAM,mahjoubfar2017time,goda2013dispersive}. Frequency-domain and inverse $4f$ approaches have also been demonstrated with extremely high effective frame rates by mapping temporal evolution into spectral or spatial frequency domains~\cite{Zhu2022FISI}.

In addition to computational approaches, fully optical single-shot techniques have been also proposed. Sequentially timed all-optical mapping photography (STAMP) maps time into wavelength via ultrafast spectral encoding~\cite{Nakagawa2014STAMP}. Single-shot non-synchronous array photography (SNAP) uses spatially distributed optical delays based on diffractive optical elements and glass echelons to capture temporally delayed replicas of a scene~\cite{Sheinman2022SNAP}. Although these techniques provide impressive performance, they typically rely on specialized components such as broadband femtosecond sources, custom-fabricated echelons, streak cameras, or ultrafast detectors, together with complex calibration procedures. As a result, the hardware cost of state-of-the-art ultrafast imaging systems typically reaches tens to hundreds of thousands of US dollars, which remains a significant barrier for widespread adoption in laboratories without dedicated ultrafast instrumentation.

In recent years, light-field and spatially multiplexed imaging approaches have been studied extensively as efficient parallel acquisition schemes that record multiple views or replicas of a scene within a single camera exposure~\cite{tian20153d,tian2014multiplexed,yu2021microlens}. Such architectures enable three-dimensional reconstruction and improved acquisition efficiency, and also suggest a promising route toward ultrafast imaging by replacing temporal sampling with parallel spatial channels. However, to the best of our knowledge, existing implementations of spatially multiplexed imaging mostly target spatial or angular information and have not been fully exploited for direct ultrafast temporal imaging based on passive optical encoding.

Here, we present a low-cost passive single-shot ultrafast imaging architecture that directly maps temporal evolution into spatially separated image replicas through controlled optical path delays. The proposed system consists of a commercial microlens array, a stack of standard microscope cover glasses, and a consumer-grade CMOS image sensor, and its total hardware cost is below US\$500. By varying the number of cover glasses in the path of each microlens channel, a unique cumulative optical delay is introduced per channel, and all delayed replicas are recorded in a single camera exposure. With this simple and low-cost design, the proposed system captures the evolution of a picosecond laser pulse with a temporal sampling interval of 1.46~ps, an effective frame rate of 685~Gfps, and a sequence depth of ten frames, while maintaining near diffraction-limited spatial resolution. The temporal fidelity and spatial resolution of the proposed architecture are experimentally characterized.

\begin{figure}[t!]
\centering
\begin{minipage}{\linewidth}
\centering
\includegraphics[width=0.95\linewidth]{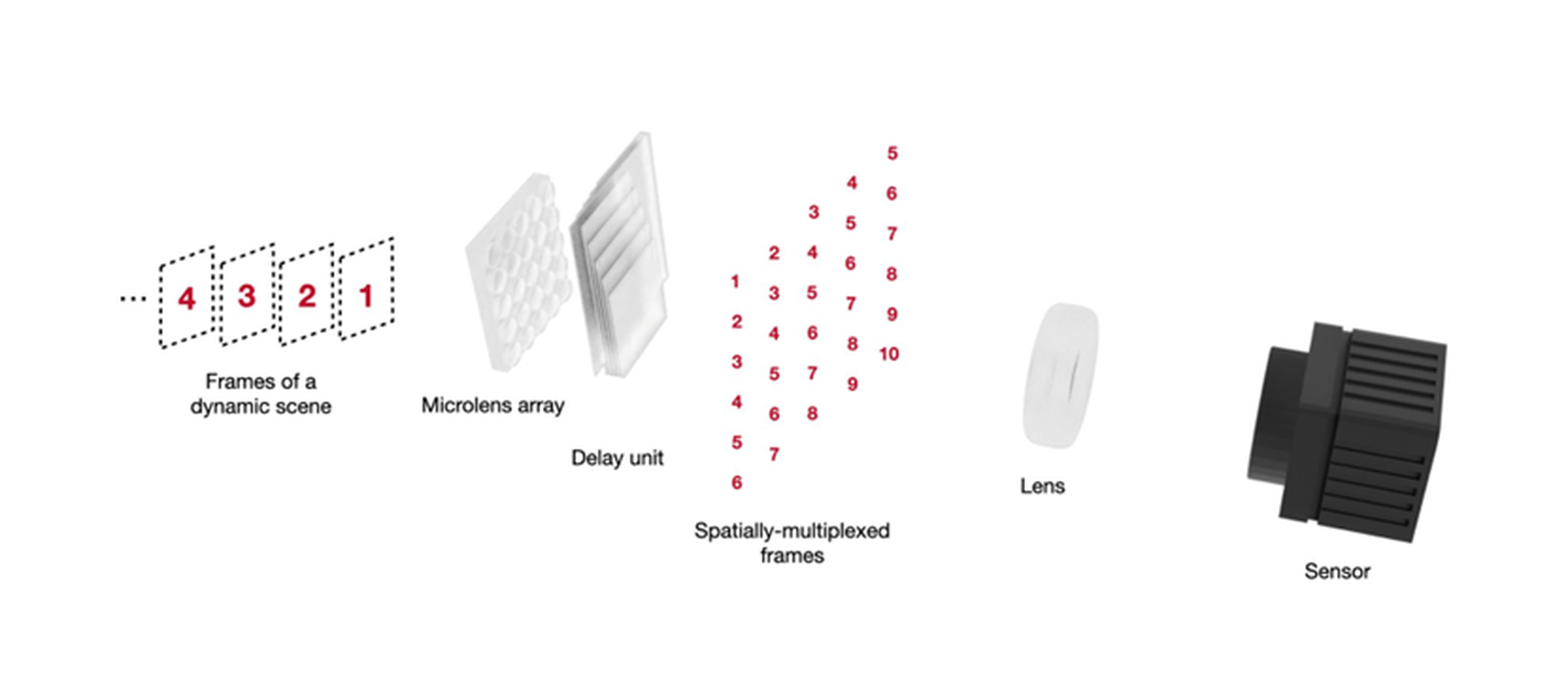}\\[-2pt]
{\small (a)}
\end{minipage}\\[4pt]
\begin{minipage}{\linewidth}
\centering
\includegraphics[width=0.75\linewidth]{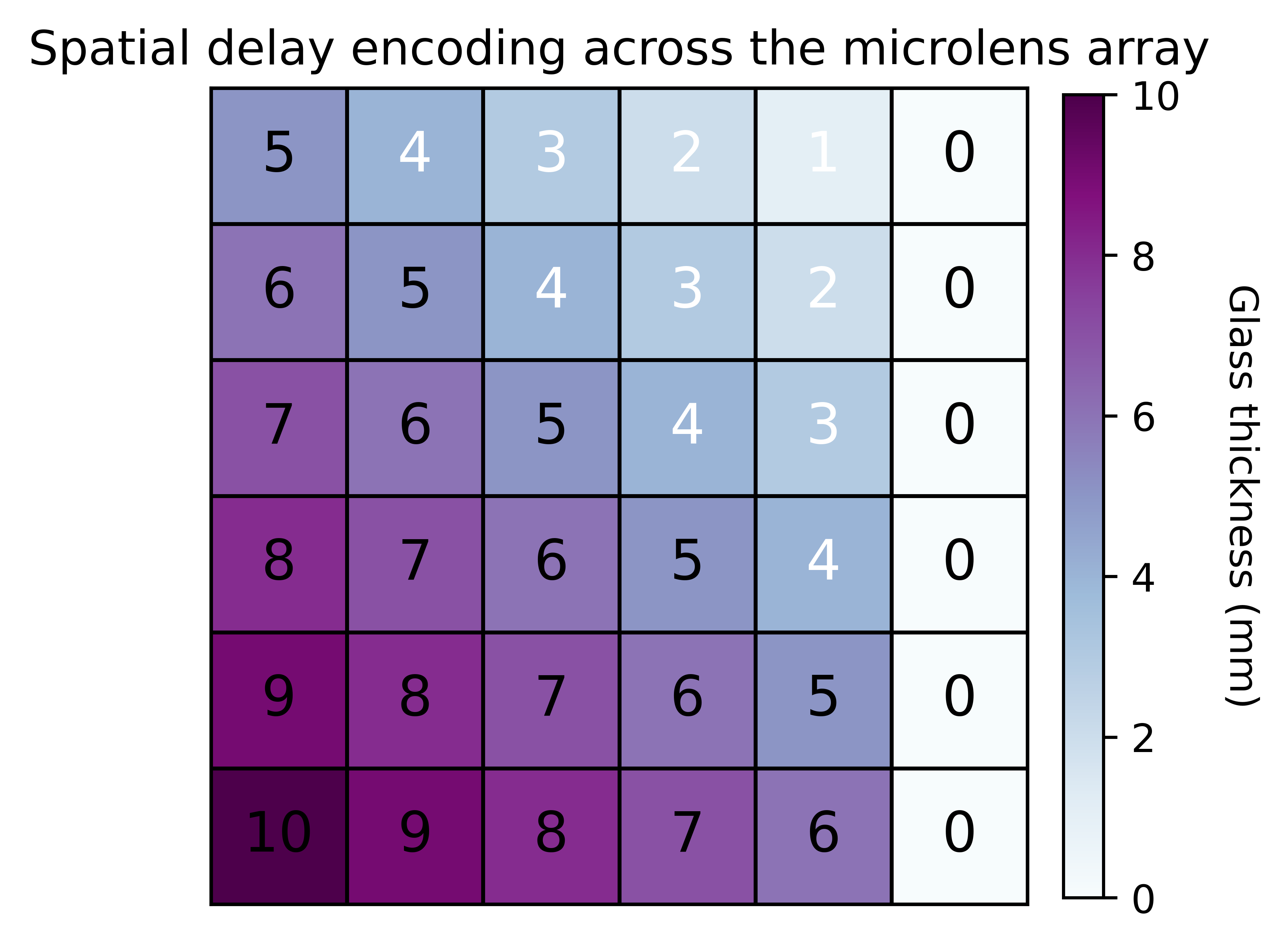}\\[-2pt]
{\small (b)}
\end{minipage}
\caption{(a) Schematic of the proposed low-cost passive spatially multiplexed ultrafast imaging system. A microlens array generates replicated image channels, each of which experiences a different optical delay introduced by a stack of standard microscope cover glasses. Temporally delayed replicas are simultaneously recorded within a single camera exposure using a consumer-grade CMOS sensor. (b) Spatial distribution of the glass thickness used for delay encoding across the $6\times 6$ microlens array. Each channel experiences a different cumulative optical delay determined by the total glass thickness along its path. The rightmost column contains no glass plates and corresponds to zero delay.}
\label{fig:1}
\end{figure}

The experimental configuration of the proposed system is illustrated in Fig.~\ref{fig:1}(a). The system consists of four main parts: image replication using a microlens array, spatially varying optical delay encoding by stacked cover glasses, relay imaging, and single-shot detection with a consumer-grade CMOS sensor.

A $6\times 6$ microlens array (MLA) is used to generate spatially separated replicas of the input scene, consistent with microlens-based light-field imaging architectures that distribute spatial information across parallel imaging channels~\cite{ng2005light,levoy2009recording}. The MLA produces 36 sub-images that are simultaneously relayed onto the camera sensor. Each sub-image corresponds to an independent spatial channel that can be assigned a unique optical delay. The spacing between adjacent sub-images is selected to prevent overlap on the sensor while maximizing the number of available temporal channels.

Temporal encoding is achieved by stacking standard microscope cover glasses directly after the MLA. By varying the total glass thickness encountered by each channel, controlled optical path-length differences are introduced across the MLA grid. The spatial distribution of the cover glass thickness used for delay encoding is shown in Fig.~\ref{fig:1}(b). The glass plates are arranged such that each channel experiences a unique cumulative delay along both the rows and columns of the MLA, producing a sequence of progressively delayed replicas spanning ten temporal frames. Because each delay element is a commodity cover glass, the cost of the temporal encoding stage is negligible compared to the detector.

Each replica produced by the MLA propagates through a distinct optical path length and arrives at the camera sensor with a relative temporal delay arising from the accumulated glass thickness~\cite{akturk2004pulse}. The delay introduced by a glass plate of thickness $d$ and refractive index $n$ is

\begin{equation}
\Delta t = \frac{(n-1)d}{c},
\end{equation}

where $c$ is the speed of light in vacuum~\cite{Goodman2005}. For the microscope cover glass used in this work ($n \approx 1.44$), a thickness increment of 1~mm introduces a temporal delay of approximately 1.46~ps.

If the delay assigned to the $(i,j)$-th channel is $\tau_{i,j}$, the camera records the exposure-integrated intensity

\begin{equation}
S_{i,j}(x,y)=
\int_{T_{\mathrm{exp}}}
I(x,y,t-\tau_{i,j})\,dt,
\end{equation}

where $T_{\mathrm{exp}}$ is the camera exposure time. Since the exposure duration greatly exceeds the temporal extent of the ultrafast event, each spatial channel records the time-integrated intensity of a temporally shifted replica of the scene. Temporal information is therefore encoded optically via relative propagation delays before detection, such that different spatial channels correspond to distinct relative times.

During a single exposure, all delayed replicas are recorded simultaneously and form a mosaic image composed of spatially separated sub-images associated with different optical delays. By segmenting the mosaic according to the microlens array grid and ordering the channels by increasing delay, a time-resolved sequence is reconstructed from a single camera exposure. The delayed replicas are imaged onto a consumer-grade CMOS camera by a relay imaging lens, and the camera records a single intensity image in which all temporally delayed replicas are captured simultaneously.

The proposed passive spatial multiplexing approach introduces an intrinsic trade-off between the temporal sampling density and the spatial resolution. For a detector with a total pixel count $P_x \times P_y$ and an MLA producing $N_x \times N_y$ spatial channels, the available detector area is partitioned among the replicated image channels, and each reconstructed sub-image contains

\begin{equation}
P_{\mathrm{sub}} =
\frac{P_x P_y}{N_x N_y},
\end{equation}

pixels, which reflects the division of detector sampling capacity among parallel spatial channels as commonly observed in multiplexed and light-field imaging architectures~\cite{ng2005light,tian20153d}. The achievable temporal window is determined by the maximum optical delay difference between the earliest and the latest spatial channels,

\begin{equation}
T = (N-1)\Delta t,
\end{equation}

where $N$ is the number of reconstructed temporal frames separated by a delay step $\Delta t$. For a temporal sampling interval of 1.46~ps and a sequence depth of ten frames, the observable temporal window spans approximately 13~ps.

Assuming diffraction-limited imaging with numerical aperture $\mathrm{NA}$ and wavelength $\lambda$, the spatial resolution within each sub-image is given by the Rayleigh criterion~\cite{born2013principles},

\begin{equation}
\delta x \approx \frac{0.61\lambda}{\mathrm{NA}},
\end{equation}

so that practical system performance is governed by both optical diffraction and spatial channel partitioning. This relationship indicates that, for passive time-to-space mapping systems, temporal scalability is ultimately constrained by the available detector area rather than the electronic bandwidth.

To experimentally validate the proposed architecture, both its temporal and spatial performance are characterized. For the temporal characterization, a single picosecond laser pulse is isolated and injected into the imaging path, similar to prior visualization of propagating optical wavefront dynamics~\cite{liang2017single}, and its transient evolution is encoded across the spatially delayed channels within a single camera exposure. After the acquisition, the mosaic image is segmented into individual sub-images, which are ordered according to their designed delays, and a time-resolved frame sequence is reconstructed.

\begin{figure}[t!]
\centering
\includegraphics[width=\linewidth]{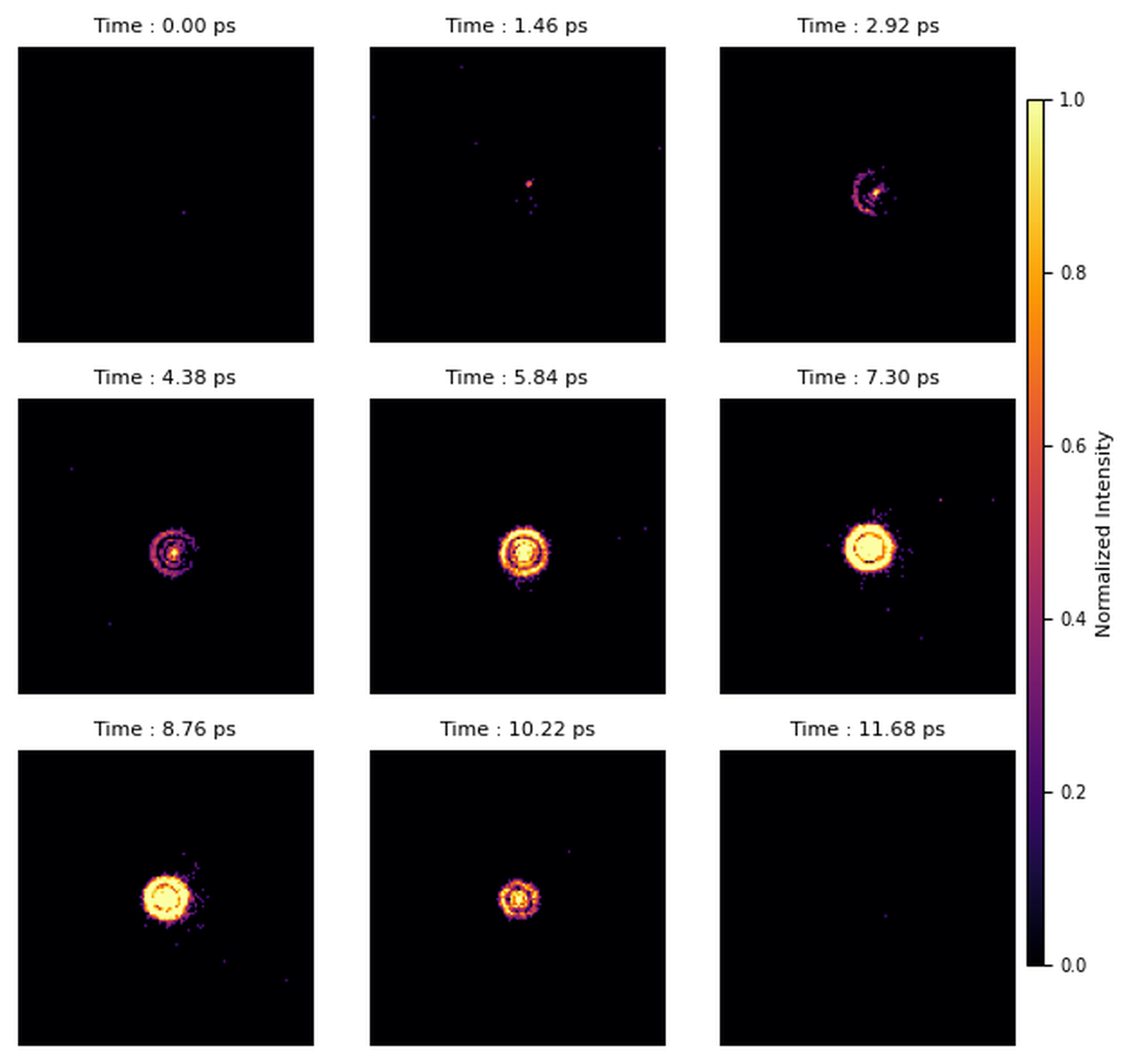}
\caption{Single-shot reconstruction of the temporal evolution of a picosecond laser pulse. Each sub-image corresponds to a different optical delay introduced by the proposed spatial multiplexing architecture. The sequence is recovered from a single camera exposure with a temporal spacing of 1.46~ps between the frames.}
\label{fig:3}
\end{figure}

\begin{figure}[t!]
\centering
\includegraphics[width=\linewidth]{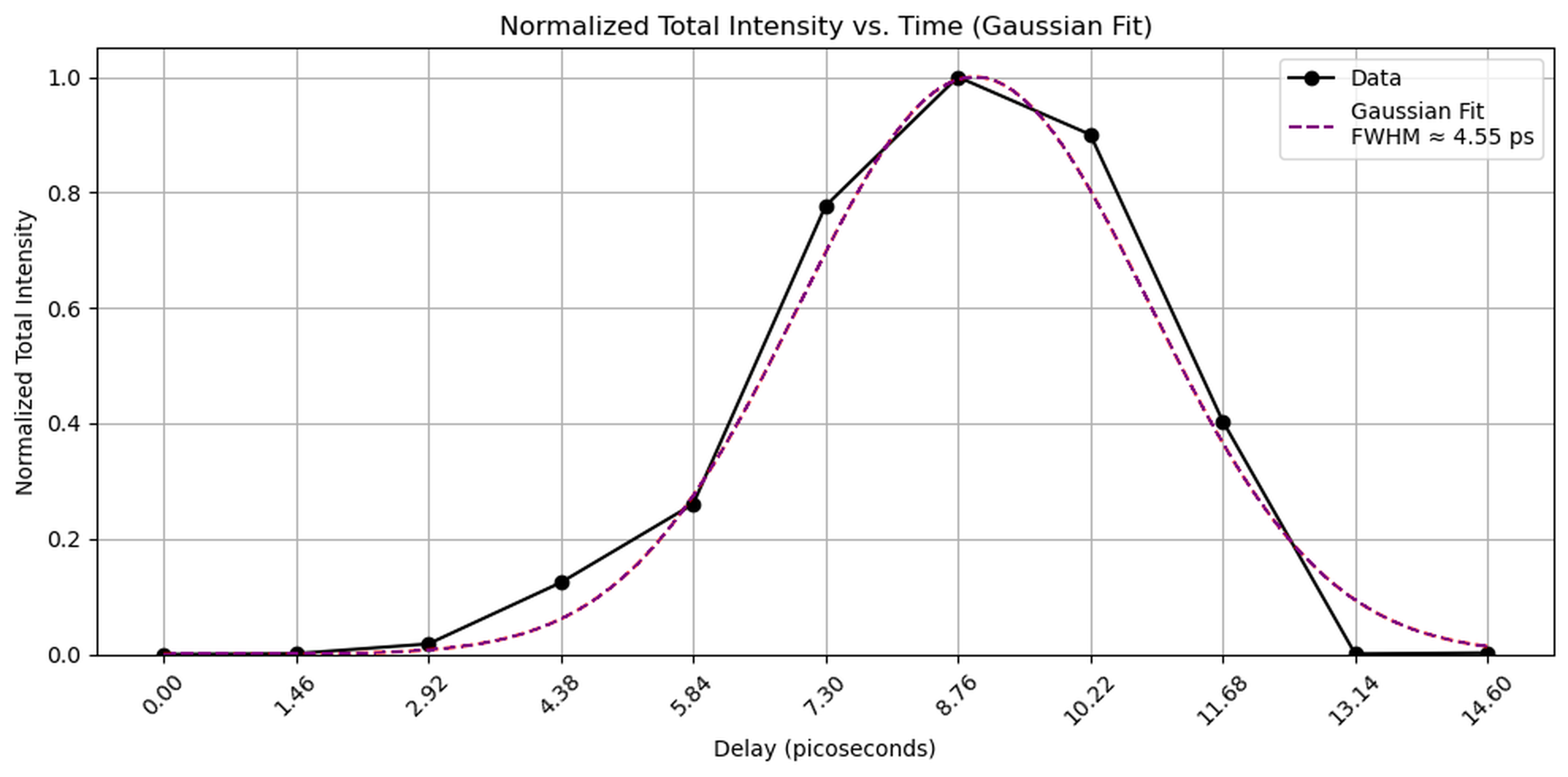}
\caption{Normalized total intensity extracted from reconstructed frames as a function of relative temporal delay. The measured temporal profile follows a Gaussian distribution (dashed curve), which confirms the accurate preservation of the pulse dynamics.}
\label{fig:4}
\end{figure}

A total of ten temporally ordered frames are obtained from a single camera exposure with a temporal spacing of 1.46~ps, as shown in Fig.~\ref{fig:3}. The reconstructed sequence clearly reveals the temporal evolution of the laser pulse intensity, and the effective frame rate of the system is therefore 685~Gfps.

To quantify the temporal fidelity of the system, the total intensity within each frame is calculated after the background subtraction and the normalization. The resulting normalized intensity values are plotted as a function of the relative delay assigned to each channel in Fig.~\ref{fig:4}. The measured temporal profile closely follows a Gaussian distribution, which is consistent with the expected shape of the laser pulse and confirms that the proposed passive spatial multiplexing and delay encoding scheme accurately preserves the temporal dynamics of the scene. The recovered pulse width also provides an independent cross-check of the 1.46~ps delay step between the channels.

\begin{figure}[t!]
\centering
\includegraphics[width=\linewidth]{fig4.png}
\caption{Spatial resolution characterization of the imaging system using a point-source measurement. The red square indicates the selected region of interest. The inset shows the background-subtracted point spread function used for Gaussian fitting and estimation of the system full width at half maximum.}
\label{fig:6}
\end{figure}

The spatial resolving capability of the proposed imaging system is next characterized using a point spread function (PSF) measurement, as shown in Fig.~\ref{fig:6}. The measured PSF exhibits FWHM values of 1.86 and 1.62 pixels along the horizontal and vertical axes, respectively, with an ellipticity of 1.15. These values indicate near diffraction-limited sub-image formation, and confirm that the proposed passive spatial multiplexing and delay encoding introduce only minor aberrations. The combined point spread function of the system arises from the microlens imaging, the relay optics, and the propagation through multiple cover glass delay elements. In addition to diffraction, residual aberrations may also arise from angular deviations introduced by thickness transitions between adjacent delay plates, and from alignment tolerances across the MLA channels.

Assuming incoherent imaging, the overall system PSF can be approximated as the convolution of the individual subsystem responses, consistent with linear shift-invariant imaging theory~\cite{Goodman2005},

\begin{equation}
\mathrm{PSF}_{\mathrm{sys}}
= \mathrm{PSF}_{\mathrm{MLA}}
\otimes
\mathrm{PSF}_{\mathrm{relay}}
\otimes
\mathrm{PSF}_{\mathrm{delay}},
\end{equation}

so that the spatial resolution degradation remains limited, provided that delay-induced wavefront distortions are small compared to the diffraction-limited spot size. The experimentally observed near-circular PSF therefore confirms that the proposed passive delay encoding does not significantly distort the imaging wavefront.

The proposed passive spatial multiplexing scheme introduces a trade-off between spatial resolution, temporal resolution, and sequence depth. Increasing the number of temporal frames requires additional spatial channels, which in turn reduces the pixel count available per sub-image. Depending on the target application, this trade-off can be optimized in practice. For example, transient waveform characterization may prioritize temporal sampling density, whereas imaging-based diagnostics may favor increased spatial resolution per channel. The passive nature of the proposed architecture enables flexible adaptation through purely geometric scaling without modification of the detection electronics, and the measured point spread function confirms that acceptable spatial resolution can be maintained even under sub-picosecond temporal sampling.

Several practical factors also influence the achievable temporal accuracy of the proposed architecture. The effective timing precision is determined not only by the nominal optical delay step but also by fabrication tolerances of the cover glass thickness, refractive-index variations, and alignment errors across the microlens array channels. Small deviations in the optical path length introduce timing jitter between the spatial channels, which may slightly broaden the temporal response. In our implementation, these effects remain significantly smaller than the designed delay increment, and therefore do not limit the frame separation. Future implementations may further improve timing uniformity by using precision-fabricated delay elements or monolithic optical substrates. The temporal resolution can also be extended toward the sub-picosecond regime by using thinner delay increments or higher-index materials while maintaining fully passive operation.

In conclusion, we experimentally demonstrate, to the best of our knowledge, the first fully passive single-shot ultrafast imaging architecture based on direct time-to-space mapping through spatial multiplexing and optical delay encoding with a total hardware cost below US\$500. The proposed system consists only of a commercial microlens array, a stack of standard microscope cover glasses, and a consumer-grade CMOS image sensor. The system captures the evolution of a picosecond laser pulse with a temporal sampling interval of 1.46~ps, an effective frame rate of 685~Gfps, and a sequence depth of ten frames. The temporal fidelity of the architecture is confirmed by the recovery of the expected Gaussian pulse profile from a single exposure, and the point spread function measurement verifies near diffraction-limited spatial performance with FWHM values of 1.86 and 1.62 pixels along the horizontal and vertical directions. Compared to streak-camera- or echelon-based ultrafast imaging systems, the proposed architecture reduces the hardware cost by approximately two to three orders of magnitude, while eliminating the need for computational reconstruction, active temporal modulation, and specialized ultrafast detectors. We believe the reported passive ultrafast imaging architecture is an attractive platform for investigating transient optical dynamics, nonlinear propagation effects, transient scattering phenomena, and light-matter interaction in laboratory environments without dedicated ultrafast instrumentation, and can be readily scaled toward higher frame rates and longer temporal windows by adjusting the MLA geometry and glass thickness increments while preserving its simple, low-cost, and fully passive design.

\begin{backmatter}

\bmsection{Funding} My article has no associated award funding.

\bmsection{Disclosures} The authors declare no conflicts of interest.

\bmsection{Data availability} Data underlying the results presented in this paper are not publicly available at this time but may be obtained from the authors upon reasonable request.

\end{backmatter}

\bibliography{references}

\bibliographyfullrefs{references}

\end{document}